\documentclass[12pt]{article}
\usepackage{amsmath}

\renewcommand{\a}{\alpha} 

\renewcommand{\d}{\delta}

\renewcommand{\l}{\lambda} 
 
\newcommand{\m}{\mu} 
\newcommand{\n}{\nu}
\newcommand{\p}{\pi} 
\renewcommand{\r}{\rho} 
\newcommand{\s}{\sigma}
\renewcommand{\S}{\Sigma}

\newcommand{\tx}{\tilde{x}}
\newcommand{\ty}{\tilde{y}}
\newcommand{\tz}{\tilde{z}}
\newcommand{\tV}{\tilde{V}}
\newcommand{\ts}{\tilde{\sigma}}

\newcommand{\cL}{{\cal L}}
\newcommand{\cP}{{\cal P}}

\newcommand{\pd}{\partial}

\newcommand{\half}{\frac{1}{2}}

\begin{document}

\author{Jiri Hoogland\footnote{jiri@cwi.nl} 
  and Dimitri Neumann\footnote{neumann@cwi.nl}\\
  CWI, P.O.~Box 94079, 1090 GB  Amsterdam, The Netherlands}
\title{\textbf{Scale invariance and contingent claim pricing}}
\maketitle

\thispagestyle{empty}
\begin{abstract}
  Prices of tradables can only be expressed relative to each other at
  any instant of time.  This fundamental fact should therefore also
  hold for contingent claims, i.e. tradable instruments, whose prices
  depend on the prices of other tradables. We show that this property
  induces local scale-invariance in the problem of pricing contingent
  claims.  Due to this symmetry we do {\it not\/} require any
  martingale techniques to arrive at the price of a claim. If the
  tradables are driven by Brownian motion, we find, in a natural way,
  that this price satisfies a PDE.  Both possess a manifest
  gauge-invariance. A unique solution can only be given when we impose
  restrictions on the drifts and volatilities of the tradables, i.e.
  the underlying market structure. We give some examples of the
  application of this PDE to the pricing of claims. In the
  Black-Scholes world we show the equivalence of our formulation with
  the standard approach. It is stressed that the formulation in terms
  of tradables leads to a significant conceptual simplification of the
  pricing-problem.
\end{abstract}

\newpage

\section{Introduction}
\label{sec:introduction}
The essence of trading is the exchange of goods.  Every transaction
sets a ratio between the value of the two goods. This means that there
is no such thing as the absolute value of an object, it can only be
defined relative to the value of another object.  If we only have one
asset, we cannot assign a price to the asset. We need at least two
assets. Then after choosing one of these two assets, the other asset
can be assigned a price relative to the first one. If we have $n+1$
tradable assets we can choose any of these $n+1$ tradables to assign a
price to the other ones. The asset that is chosen to set the prices of
the other asset is often called a numeraire. In fact, we have even
more freedom. We can choose any positive-definitive function as a
numeraire and express every asset price in terms of it, e.g. money.

Thus a price is always given in terms of some unit of measurement.  It
is a measure-stick which is used to relate different objects. As long
as everything is expressed in terms of this one unit prices can be
compared. Whether we scale the unit does not matter, prices will scale
accordingly. This scale-invariance is of great importance. Not only
the prices of tradables which are used to set up the basic economy
should scale with a change in numeraire, but any derived tradable like
contingent claims, depending on other tradables, should act in the
same way. This leads in a natural way to the constraint that the price
of a claim as a function of the underlying tradables should be
homogeneous\footnote{A function $f(x_1,\ldots,x_n)$ is called
  homogeneous of degree $\r$ if $f(ax_1,\ldots,ax_n)=a^\r
  f(x_1,\ldots,x_n)$.  Homogeneous functions of degree $\r$ satisfy
  the following property (Euler): $\sum_{i=1}^nx_i\frac{\pd}{\pd
    x_i}f(x_1,\ldots,x_n)=\r f(x_1,\ldots,x_n)$} of degree $1$.
Otherwise the economy is not well posed.

Although Merton~\cite{Merton73} already noticed the homogeneity
property for the case of a simple European warrant, it was apparently
not recognized that this property should be an intrinsic property of
any economy in which tradables and derivatives on these tradables have
prices relative to some numeraire. More recently,
Jamshidian~\cite{Jamshidian97} discussed interest-rate models and
showed that if a payoff is a homogeneous function of degree $1$ in the
tradables, it leads naturally to self-financing trading strategies for
interest-rate contingent claims.  But again it is not appreciated that
the homogeneity is a fundamental property, which any economy should
possess to be properly defined.

To compute the price of a contingent claim~\cite{HarrisonPliska81} one
normally starts with the definition of the stochastic dynamics of the
underlying tradables.  The next step is to find a self-financing
trading strategy which replicates the payoff of the claim at the
maturity of the contract. If the economy does not allow for arbitrage
and is complete, this self-financing trading strategy gives a unique
price for the claim price. To arrive at this result, one has to find a
measure under which the tradables, discounted by a numeraire, are
martingales. This requires a change of measure. When this change of
measure exists, we have to show that the discounted payoff of the
claim is a martingale under this new measure too. Then the martingale
representation theorem is invoked to link the discounted payoff
martingale to the underlying discounted tradables.  This then gives a
self-financing trading strategy using underlying tradables, which
replicates the claim at all times and thus yields a price for the
claim.  The invariance of the choice of numeraire is reflected in the
fact that the price of the claim is indeed invariant under changes of
measure, which are associated with different numeraires. Geman {\em
  et.al.\/}~\cite{GemanElkarouiRochet95} used this invariance to show
that, depending on the pricing problem at hand, it is useful to select
a numeraire, which most naturally fits the payoff of the claim.

In this paper we start our discussion with the scale-invariance of a
frictionless economy of tradables with prices expressed in an
arbitrary numeraire. We assume the economy to be complete.  Our next
step is to define the stochastic dynamics of the prices of tradables.
It\^o then leads to a SDE for a claim-price. If the claim-price solves
a certain PDE then together with the homogeneity property this leads
automatically to a self-financing trading strategy replicating the
claim price. If no-arbitrage constraints are imposed on the drifts and
volatilities of the stochastic prices, this price is unique.  The
invariance under changes of numeraire becomes very transparent due to
the homogeneity-property. We do not have to apply changes of measure
and this leads in our view to a conceptually more satisfying and
transparent contingent claim pricing argument.  Finally the
scale-invariance property should be satisfied also in economies which
do have friction. The symmetry invokes constraints which may be useful
in model-building, e.g. more general stochastic processes.  We will
discuss this in a forthcoming publication~\cite{HooglandNeumann99b}.
Also a more rigorous exposition of these results will be presented in
this publication. In the present paper, we want to focus on the
main ideas and defer the mathematical details to a later time.  To the
best of our knowledge this is the first time that the consequences of
the scale-invariant economy for contingent-claim pricing have been
outlined and discussed.

The outline of the article is as follows.  In
section~\ref{sec:contingent-claim-pricing} we introduce some standard
notions used to price contingent claims in an economy with stochastic
tradables. In subsection~\ref{sec:homogeneity} we show that for an
economy to be properly defined it is required to be scale-invariant.
The scaling-symmetry restricts the contingent claim price: it should
be a homogeneous function of the underlying tradables of degree 1.  In
subsection~\ref{sec:dynamics-market-model} we introduce the dynamics
of the prices of tradables and introduce the notion of deterministic
constraints on the dynamics, which may follow from certain choices for
the drifts and volatilities of the tradables. In
subsection~\ref{sec:deriving-basic-pde} we use the homogeneity
together with It\^o to derive a PDE for the contingent claim value.
The homogeneity automatically insures the existence of a
self-financing trading strategy for the contingent-claim. In
subsection~\ref{sec:uniq-arbitr-revis} we show that the claim price
will be unique if the constraints on the dynamics can be written as
self-financing portfolios. Finally in
subsection~\ref{sec:gauge-invariance-pde} it is shown that the
symmetry is inherited by the PDE for the claim value. This allows us
to pick an appropriate numeraire (fix a gauge) and solve the PDE.
Section~\ref{sec:applications} gives various applications of the PDE
and the scale-invariance in pricing of contingent claims. In
subsection~\ref{sec:gener-solut-logn} we give the explicit formula for
a European claim with log-normal prices for the underlying tradables.
In subsection~\ref{sec:black-scholes} it is shown that the
Black-Scholes PDE is contained in our approach. In
subsection~\ref{sec:quantos} the pricing of quantos is discussed. In
our formulation the pricing becomes trivial. In
subsection~\ref{sec:HJM} we show that term-structure models fit
naturally into our approach and give as an example the price of a
log-normal stock in a gaussian HJM model. Another example of the
simple formulae is given in subsection~\ref{sec:trigger-swap}, where
we consider a trigger-swap.  Finally we give our conclusions and
outlook in section~\ref{sec:conclusions-outlook}.

\section{Contingent claim pricing}
\label{sec:contingent-claim-pricing}
In the following subsections we will discuss some general properties of
contingent claim pricing using dimensional analysis.

First let us recall the basic principles.  We consider a frictionless
market with $n+1$ tradables\footnote{We will always use Greek symbols
  for indices running from $0$ to $n$ and Latin symbols for indices
  running from $1$ to $n$. Furthermore, we use Einstein's summation
  convention: repeated indices in products are summed over.} with
prices $x_\m$, where $\m=0,\ldots,n$. The prices
$x\equiv\{x_\m\}_{\m=0}^n$ follow stochastic processes, driven by
Brownian motions\footnote{More general processes will be discussed in
  Ref.~\cite{HooglandNeumann99b}}. Time is continuous.  Transaction
costs are zero.  Dividends are zero.  Short positions in tradables are
allowed.  We want to value a European claim at time $t$ promising a
payoff $f(x)$ at maturity $T>t$. To attach a rational price to the
claim at time $t$ we have to find a dynamic portfolio or trading
strategy $\phi\equiv\{\phi_\m(x,t)\}_{\m=0}^n$ of underlying tradables
$x$ with value
$$
V(x,t)=\phi_\m(x,t)x_\m
$$
which replicates the payoff of the claim at maturity, $V(x,T)=f(x)$.
Let us apply It\^o to the trading strategy:
$$
dV=\phi_\m dx_\m+x_\m d\phi_\m+d[\phi_\m,x_\m]
$$
Here $[\phi_\m,x_\m]$ stands for the quadratic variation\footnote{Or
  covariance.} of the two processes.  We assume that the $\phi$ are
adapted to $x$, predictable, i.e.  given the values of $x$ up to time
$t$ we know the $\phi$.  This implies
$$
d[\phi_\m,x_\m]=0
$$
Furthermore the trading-strategy has to be self-financing, i.e. we set
up a portfolio for a certain amount of money today such that no
further external cash-flows are required during the life-time of the
contract to finance the payoff of the claim at maturity.  All changes
in the positions $\phi_\m(x,t)$ at any given instant are financed by
exchanging part of the tradables at current market prices for others
such that the total cost is null:
$$
x_\m d\phi_\m=0
$$
If we can find such a trading-strategy, then the rational value of the
claim today equals the value of the trading portfolio today.  If there
is a non self-financing trading-strategy, the claim value at time $t$
will not be unique. Hence arbitrage opportunities exist.  Uniqueness
of the claim value only follows in special cases, i.e. for specific
choices of stochastic dynamics and drifts and volatilities.  This will
be discussed in more detail in Sec.~\ref{sec:uniq-arbitr-revis}.
The self-financing property of the trading-strategy is expressed as
follows.
$$
d V=\phi_\m dx_\m
$$
Finally we also have to impose the following restriction on the
allowed tra\-ding\-strategies $\phi$ to be admissible: the value of a
self-financing replicating portfolio is either deterministically zero
at any time during the life of the contract or never. Otherwise
arbitrage is possible. We come back to this point in
Sec.~\ref{sec:uniq-arbitr-revis}.

\subsection{Homogeneity}
\label{sec:homogeneity}
For a market to exist we need at least two tradables. Prices are
always expressed in terms of a numeraire. The numeraire may be any
positive-definite, possibly stochastic, function. The freedom to
choose an arbitrary numeraire implies the existence of a
scaling-symmetry for prices.  The symmetry automatically implies the
existence of a delta-hedging strategy for any tradable which depends
on other underlying tradables.

Let us consider again a market with $n+1$ basic tradables with prices
$x$ at time $t$.  These prices are in units $U$ of the numeraire.  We
say that the $x$ have dimension $U$, or symbolically $[x_\m]=U$.  For
the moment we leave the dynamics unspecified.  What can be said about
the price of a claim today, again in units of $U$, when expressed in
terms of the tradables $x$?  Let us denote the price of the claim by
$V(x,t)$. Just on the basis of dimensional analysis we can write down
the following form for the price
\begin{equation}
\label{eq:1}
V(x,t)=\phi_\m(x,t)x_\m 
\end{equation}
Since $[V]=U$ and $[x_\m]=U$, the functions $\phi_\m$ are
dimensionless, $[\phi_\m]=1$. This implies that they can only be
functions of ratios of different tradables, which are again
dimensionless.

The same arguments apply to any payoff function, for else it is
ill-specified. For example, the payoff-function of a vanilla call with
maturity $T$ does not seem to have this form at first sight
$$
(S(T)-K)^+
$$
But what is meant is the following function of a stock $S(t)$ and a
discount bond $P(t,T)$, which pays $1$ unit of $U$ at time $T$
$$
(S(T)-KP(T,T))^+
$$
and this does have the right form.

Now suppose that we change our unit of measurement. If we scale the
unit by $a$, such that $U\to U/a$, then the prices of the tradables
will scale accordingly, $x_\m\to a x_\m$. Using the dimensional
analysis result above we then find the following property for the
price of the claim
\begin{equation}
  \label{eq:2}
  V(a x,t)
  =\phi_\m(a x,t) a x_\m
  =a \phi_\m(x,t) x_\m
  =a V(x,t)
\end{equation}
The price of the claim is a homogeneous function of degree $1$.  Note
the scaling factor $a$ may be local, $a=a(x,t)$. Differentiating
Eq.~\ref{eq:2} with respect to $a$, this immediately yields the
following relation, valid for any homogeneous function\footnote{We
  allow generalized functions.} of degree $1$,
\begin{equation}
  \label{eq:3}
  V(x,t)
  =\frac{\pd V(x,t)}{\pd x_\mu} x_\m
  \equiv V_{x_\m}(x,t) x_\m
\end{equation}
This result is independent of the choice of dynamics.  Even if we
relax the frictionless market assumptions, this scaling-symmetry
should not be broken.

As already mentioned various authors~\cite{Merton73,Jamshidian97}
already touched upon the homogeneity-property of certain claim prices,
but they always inferred this property as a consequence of the
no-arbitrage conditions they imposed on the drift and volatilities of
the tradables. Furthermore their claim is that this property only
holds in certain cases. In fact Jamshidian ~\cite{Jamshidian97} gives
a theorem which is very similar to what we discuss in
subsection~\ref{sec:deriving-basic-pde}, except that he doesn't
recognize the fact that the required homogeneity should always be
satisfied.  This should be contrasted with our presentation above,
where we show that this homogeneity property is one of the most
fundamental properties any market model must posses to be well-posed.
The homogeneity property just expresses the fact that one needs a
proper coordinate-system.  It could be termed: `the relativity
principle of finance'.
 
\subsection{Dynamics: the market model}
\label{sec:dynamics-market-model}
The prices of tradables, relative to a numeraire, change over time.
Let us assume that the dynamics of the tradables is given by the
following stochastic differential equation:
\begin{equation}
  \label{eq:4}
  d x_\m(t)=\a_\m(x,t)x_\m(t)dt+\s_\m(x,t)x_\m(t)\cdot dW(t)
\end{equation}
where we have $k$ independent Brownian motions driving the $n$
tradables and initial conditions\footnote{Here $\s_\m$ and $dW$ should
  be understood as $k$-dimensional vectors. We denote the inner
  product by a dot.} $x_\m(t)$. The Brownian motion is defined under
the measure with respect to the numeraire. This is often called the
real-world measure in the literature. To determine a price for the
claim we will {\em always\/} work under this measure. This should be
contrasted with the usual approach, where one first applies a change
of measure to make the tradables martingales under the new measure.
Then one invokes the martingale representation theorem to determine
the claim price. This change of measure is not required, as we will
show later, for the determination of a rational price. In fact we do
not even have to require the tradables to be strictly positive.  If
one of the tradables would become zero, this is allowed as long as it
hits zero in a non-deterministic way. The tradable should not be used
as a numeraire.

For the properties of the drift and volatilities we refer to
Appendix~\ref{sec:stoch-diff-equat}. It is convenient to extract a
unit of $x_\m$ from the drift and volatilities to make the LHS of
Eq.~\ref{eq:4} dimensionless.  Then the RHS should be a homogeneous
function\footnote{In the literature the $\a_\m$ and $\s_\m$ are often
  called {\em relative\/} drift and volatilities.} of the tradables of
degree $0$ too.  Thus the only allowed form for the drift and
volatility-structure are functions of the ratios of the tradables.
This is a fundamental requirement for any viable and properly posed
market model.

A priori it could well be that deterministic relations exist between
the tradables.  These relations should satisfy certain constraints in
order to attach a unique rational price to a claim.  If these
constraints are satisfied, arbitrage is not possible.
We will come back to this point in section~\ref{sec:uniq-arbitr-revis}.

\subsection{Deriving the basic PDE}
\label{sec:deriving-basic-pde}
The results of the previous sections are precisely what is needed to
obtain a PDE for the price of a contingent claim. It will be shown
that the homogeneity-property, together with this PDE, is all that is
necessary to obtain a unique self-financing trading-strategy in an
arbitrage-free market. We do not have to make a detour using
martingale techniques to prove this fact. This is a substantial
conceptual simplification of the standard theory.

Let us consider the evolution of the contingent claim price $V(x,t)$
in time. Using It\^o we arrive at the following SDE
$$
d V 
=\left(V_t+\half \s_\m\cdot\s_\n x_\m x_\n V_{x_\m x_\n} \right)dt
+V_{x_\m} d x_\m
$$
At this point the homogeneity property of $V(x,t)$ is used. Since 
$$
V=V_{x_\m} x_\m 
$$
we see that if the claim value solves the PDE
\begin{equation}
  \label{eq:5}
  V_t+\half \s_\m\cdot\s_\n x_\m x_\n V_{x_\m x_\n}\equiv
  \cL V=0
\end{equation}
a replicating portfolio, containing $V_{x_\m}$ of tradable $x_\m$, is
indeed self-financing.
$$
d V=V_{x_\m}d x_\m
$$
As usual, the payoff of the claim is specified as the boundary
condition of the PDE.

Note that the drift terms did not enter the derivation of the PDE at
all. We did not have to apply a change of measure to obtain an
equivalent martingale measure and use the martingale representation
theorem. All that is needed is the homogeneity of the contingent claim
price as a function of the underlying tradables.

The PDE in Eq.~\ref{eq:5} provides, in our view, the most natural
formulation of the valuation of claims on tradables in a Brownian
motion setting. It allows us to easily derive the classical result of
Black, Scholes, and Merton (subsection~\ref{sec:black-scholes}), but also
the results of Heath-Jarrow-Morton (subsection~\ref{sec:HJM}). Although we
considered European claims up till now, it is not too difficult to
include path-dependent properties.  This will be discussed in
Ref.~\cite{HooglandNeumann99b}.

\subsection{Uniqueness: No arbitrage revisited}
\label{sec:uniq-arbitr-revis}
In the previous section we showed that if the claim-value solves
Eq.~\ref{eq:5} then the replicating portfolio for the claim is
self-financing. If deterministic relations between tradables exist,
this is too strong a condition. In that case the constraints introduce
a redundancy (gauge-freedom) in the space of tradables. This implies
that we only have to solve $\cL V=0$ modulo the constraints.  The
deterministic relations between tradables allow the construction of
deterministic portfolios with zero value for all times. We will call
them null-portfolios.  Suppose that there exist $m$ deterministic
relations
$$
P_i(t)=\psi_{i,\m}(x,t) x_\m=0
$$
with $i=1,\ldots,m$. We will assume for the moment that these
relations are independent such that they span the null-space $\cP$.
Otherwise we can find a smaller set of independent constraints to span
the null-space. We also assume that the dimension of the null-space is
constant over time. Thus we can write the null-space $\cP$ as follows.
$$
\cP=\{f_i(x,t)P_i(t)|\mathrm{arbitrary } f_i(x,t)\}
$$
where the $f_i$ are predictable homogeneous functions of degree 0
w.r.t. the prices. Taking into account the constraints we require
$$
\cL V\approx0
$$
Here we use the notation $\approx0$ to write $\cL V=0$ modulo elements
in the null-space $\cP$.

The null-portfolios are either self-financing or not. In the first
case, the price of the claim is unique up to arbitrary null-portfolios
for all times. No external cash-flows are required to keep the
null-portfolio null. In the second case we can find two portfolios
which replicate the payoff at maturity but whose values diverge as one
moves away from maturity. There will be no unique price and arbitrage
is possible.

A market will have self-financing null-portfolios if the drift and
volatilities satisfy certain constraints. A null-portfolio $P=\psi_\m
x_\m\in\cP$ satisfies by definition
\begin{equation}
  \label{eq:6}
  dP\approx0
\end{equation}
Since the null-portfolio is by definition deterministic, this leads
automatically to the following constraints
\begin{equation}
  \label{eq:7}
  \frac{\pd P}{\pd x_\m}\s_\m x_\m
  =\psi_\m\s_\m x_\m+\frac{\pd\psi_\n}{\pd x_\m}\s_\m x_\m x_\n
  \approx0
\end{equation}
If a null-portfolio is self-financing, we have
$$
dP=\psi_\m dx_\m
$$
But Eq.~\ref{eq:7} immediately gives 
\begin{equation}
  \label{eq:8}
  \psi_\mu dx_\m\approx0
\end{equation}
which implies
\begin{eqnarray*}
  \psi_\m \a_\m x_\m&\approx& 0 \\
  \psi_\m \s_\m x_\m&\approx& 0
\end{eqnarray*}
If these constraints are satisfied for all null-portfolios, then the
null-portfolios will be self-financing and hence no arbitrage is
possible.

As a simple example of such constraints, let us consider two tradables
$x_{1,2}$ with one Brownian motion
$$
\frac{d x_{1,2}}{x_{1,2}}=\a_{1,2}dt+\s_{1,2} dW(t)
$$
and constant drift $\a_{1,2}$ and equal volatility $\s_{1,2}$ and initial
values $x_{1,2}(0)=1$. Note that this is the usual setting of Black-Scholes.
The SDE for the ratio $x_2/x_1$ then becomes
$$
\frac{d x_2/x_1}{x_2/x_1}
=(\a_2-\a_1-\s_1(\s_2-\s_1))dt+(\s_2-\s_1)dW
$$
If the tradables satisfy a deterministic relation, we see that this is
only possible if the volatilities are equal, $\s_1=\s_2\equiv\s$. In
that case the above SDE reduces to an ODE
$$
\frac{d x_2/x_1}{x_2/x_1}
=(\a_2-\a_1)dt
$$
Solving the ODE, we find the following deterministic relation
\begin{equation}
  \label{eq:9}
  x_2(t)=x_1(t)e^{(\a_2-\a_1)t}
\end{equation}
The existence of this relation allows us to construct a null-portfolio
with zero value and previsible coefficients for all times. Indeed
$$
P(t)=x_2(t)-x_1(t)e^{(\a_2-\a_1)t}
$$
is trivially zero.  Two cases can be distinguished. The portfolio $P$
is self-financing or it is not. Consider the evolution of $P$
$$
d P=dx_1-e^{(\a_2-\a_1)t}dx_2+(\a_2-\a_1)e^{(\a_2-\a_1)t}x_1 dt
$$
It should be clear that only if $\a_1=\a_2$ the portfolio $P$ will be
self-financing and $x_1$ can be hedged using $x_2$. Otherwise
arbitrage is possible. Intuitively this should be obvious, two
tradables with equal risk $\s$ should yield the same returns $\a$.

Let us consider the consequences for the price $V$ of a claim if
$\a_1\ne\a_2$. We construct a
portfolio $P$ with constant coefficients $\psi_{1,2}$
$$
P(t)=\psi_1 x_1(t)+\psi_2 x_2(t)
$$
If we set
$$
\psi_2=-\psi_1 e^{(a_1-\a_2)T}
$$
then the value of the portfolio at time $T$ is $P(T)=0$. However at
$t<T$ we have
$$
P(t)=\psi_1x_1(t)\left(1-e^{(a_1-\a_2)(T-t)}\right)
$$
Since $\psi_1$ can take any value, the value of the contract which
pays zero at time $T$ can have any value. But this implies that we can
ask any price $V(t)+P(t)$ for a claim paying $V(T)$ by adding an
arbitrary portfolio with $P(T)=0$.

\subsection{Gauge invariance of the PDE}
\label{sec:gauge-invariance-pde}

It was shown that a fundamental property of any viable market-model is
the scale-invariance of the prices of tradables as expressed through
the freedom of choice of the numeraire. It leads automatically to the
requirement that the claim-price should be a homogeneous function of
degree 1 in terms of prices of tradables. This invariance should be
inherited by the dynamical equations governing the price-process for
the claim. Indeed, by differentiating Eq.~\ref{eq:3} again we obtain
\begin{equation}
  \label{eq:10}
  x_\m V_{x_\m x_\n}=0
\end{equation}
Using this result it is a simple exercise to show that $\cL V$ is
invariant under the (simultaneous) substitutions
$$
\s_\m(x,t) \rightarrow \s_\m(x,t)-\l(x,t)
$$
This invariance-property represents the fact that volatility is a
relative concept. It can only be measured with respect to some
numeraire. Prices should not depend on this\footnote{This is called a
  gauge-invariance in physics' parlance and change of numeraire in
  finance parlance.}. We can exploit this freedom to reduce the
dimension of the problem.  For example, choosing $x_0$ as a numeraire
corresponds to taking $\l(x,t)=\s_0(x,t)$. Then
\begin{equation}
 \label{eq:11}
  V_t+\frac{1}{2}(\s_i(x,t)-\s_0(x,t))
    \cdot(\s_j(x,t)-\s_0(x,t))
    x_i x_j V_{x_i x_j} = 0
\end{equation}
Now one can introduce
\begin{equation}
  V(x_0,\ldots,x_n,t)=x_0 E\left(
  \frac{x_1}{x_0},\ldots,\frac{x_n}{x_0},t\right)
\end{equation}
Then $E(x_1,\ldots,x_n,t)$ again satisfies Eq.~\ref{eq:11}.
Interesting things happen when $V$ is independent of $x_0$. In that
case, $E$ is homogeneous again, the $\s_0(x,t)$ dependence drops out,
and the game can be repeated.  Furthermore it should be noted, that
the numeraire does not have to be a tradable. As stated earlier it may
be be any positive-definite stochastic function. This freedom can be
exploited to simplify calculations.  Finally recall Eqs.~\ref{eq:3}
and \ref{eq:10}. These relations give some interesting relations
between the various greeks. This can be of use in numerical schemes to
solve the PDE.

\section{Applications}
\label{sec:applications}

In this section we give several examples, which show the simplicity
and clarity with which one derives results for contingent claim prices
using the scale-invariance of the PDE.

\subsection{General solution for the log-normal case}
\label{sec:gener-solut-logn}
We compute the claim price for a path-independent European claim with
an arbitrary number of underlying tradables, when the prices of the
tradables are log-normally distributed,
$$
\frac{d x_\m(t)}{x_\m(t)}=\a_\m(t)dt+\s_\m(t)\cdot dW(t)
$$
It is easy to write the general solution for a path-independent
European claim in this case. First we perform a change of variables
$$
x_{\m} = \exp(y_{\m})
$$
such that the PDE becomes
$$
V_t+\frac{1}{2}\s_{\m}(t)\cdot\s_{\n}(t)
(V_{y_\m y_\n}-\delta_{\m\n}V_{y_\m})=0
$$
A Fourier transformation yields an ODE in $t$
$$
\tV_t-\frac{1}{2}\s_{\m}(t)\cdot\s_{\n}(t)
(\ty_{\m}\ty_{\n}-i\delta_{\m\n}\ty_{\m})
\tV=0
$$
where $i$ denotes the imaginary unit. The ODE has the solution
$$
\tV(t)=\tV(T)
  \exp\left(-\frac{1}{2}\S_{\m\n}
  (\ty_{\m}\ty_{\n}
  -i\delta_{\m\n}\ty_{\m})\right)
$$
with
$$
\S_{\m\n}\equiv\int_t^T \s_{\m}(u)\cdot\s_{\n}(u)du
$$
Since $\Sigma$ is a non-negative symmetric matrix, it can be
diagonalized as
$$
\S_{\m\n} = A_{\m\s} A_{\n\r} B_{\s\r}, \hspace{1cm}
B=\mbox{diag}(\l_0,\ldots,\l_{m-1},0,\ldots)
$$
where $A$ is an orthogonal matrix and $m$ equals the rank of $\S$ (so
$\l_i>0$ for $0\leq i<m$).  It will turn out to be convenient to
introduce the matrix
$$
\Theta_{\m\n} = \left\{ \begin{array}{ll}
A_{\m\n} \sqrt{\l_\n} & \mbox{for $\n<m$} \\
A_{\m\n}              & \mbox{otherwise}
\end{array} \right.
$$
Clearly, this matrix is invertible,
$\det\Theta=\sqrt{\l_0\cdots\l_{m-1}}$, and it satisfies
$$
\S_{\m\n} = \Theta_{\m\s} \Theta_{\n\r} \Lambda_{\s\r},
\hspace{1cm} \Lambda=\mbox{diag}(
\underset{m}{\underbrace{1,\ldots,1}},0,\ldots)
$$
We now perform an inverse Fourier transformation on the solution of
the ODE, and find
\begin{eqnarray*}
  \lefteqn{V(x_0,\ldots,x_n,t) = \frac{1}{(2\p)^{n+1}}
  \iint V(\exp(y_0),\ldots,T) }\\
  &&\times \exp\left(-\frac{1}{2}\S_{\m\n}
  (\ty_{\m}\ty_{\n}
  -i\delta_{\m\n}\ty_{\m})
  +i\ty_\m(y_\m-\ln x_\m) \right)dy d\ty\\
&=& \frac{1}{(2\p)^{n+1}}
  \iint V(x_0\exp(y_0-\frac{1}{2}\S_{00}),\ldots,T)\\
  &&\times\exp\left(-\frac{1}{2}\S_{\m\n}
  \ty_{\m}\ty_{\n}
  +i\ty_\m y_\m \right)dy d\ty
\end{eqnarray*}
Next we introduce new variables as follows
$$
  y_\m=\Theta_{\m\n} z_\n, \hspace{1cm}
  \Theta_{\m\n}\ty _\m=\tz_\n
$$
In terms of these variables, the integral becomes (note that the
Jacobian of this transformation exactly equals one)
$$
\frac{1}{(2\p)^{n+1}}
  \iint V(x_0\exp(\Theta_{0\n}z_\n-\frac{1}{2}\S_{00}),\ldots,T)
  \exp\left(-\frac{1}{2}\Lambda_{\m\n}
  \tz_{\m}\tz_{\n}
  +i\tz_\m z_\m \right)dz d\tz
$$
The integral over the $\tz_\m$ can be calculated explicitly. It gives
rise to an $m$-dimensional standard normal PDF, multiplied by some
$\d$-functions
$$
\frac{1}{(2\p)^{n+1}}
  \int \exp\left(-\frac{1}{2}\Lambda_{\m\n}
  \tz_{\m}\tz_{\n}
  +i\tz_\m z_\m \right) d\tz=
  \phi(z) \d(z_m)\cdots\d(z_n)
$$
$$
\phi(z)=\frac{1}{\left(\sqrt{2\pi}\right)^m}
\exp\left( -\frac{1}{2} \sum_{i=0}^{m-1} z^2_i \right)
$$
The integrals over $z_\m$ for $\m\geq m$ are now trivial.
To express the result in a compact form, it is useful to
introduce a set of $m$-dimensional vectors
$$
(\theta_\m)_i=\Theta_{\m i}, \hspace{1cm} 0\leq i<m
$$
These vectors in fact define a Cholesky-decomposition
of the covariance matrix. Indeed, they satisfy
$$
  \theta_\m \cdot \theta_\n = \Sigma_{\m\n}
$$
Here the inner product is understood to be $m$
dimensional. Combining all, the solution becomes
\begin{equation}
\label{eq:sol}
V(x_0,\ldots,x_n,t)=
\int \phi(z)
V(x_0\exp(\theta_0\cdot z
-\frac{1}{2}\theta_0\cdot\theta_0),\ldots,T)d^mz
\end{equation}
For homogeneous $V$, the result can be expressed in an
even more compact form
$$
  V(x_0,\ldots,x_n,t)=\int V(x_0 \phi(z-\theta_0),\ldots,
   x_n \phi(z-\theta_n),T)d^mz
$$
If the number of tradables is small we may be able to compute
Eq.~\ref{eq:sol} analytically. Otherwise we have to use numerical
techniques.

\vspace{1\baselineskip} \noindent
At this point let us remind the reader that it is
easy to include stocks in the model with known
future dividend yields. This can be done as follows.
Suppose we want to price a European claim $V$, whose
price depends on a dividend paying stock $S$. The
dividend payments occur at times $t_i$, $1\leq i\leq n$
during the lifetime of the claim. These dividends
are given as a fraction $\d_i$ of the stock-price $S(t_i)$.
The effect of the dividend payments on the
price of the claim can be incorporated by making the
substitution
$$
  S(t) \rightarrow S(t) \prod_{i=1}^n (1+\d_i)^{-1}
$$
in the price function of a similar claim, but depending
on a non dividend paying stock. Indeed, a dividend payment
at time $t_i$ has the effect of reducing the stock-price by
a factor $(1+\d_i)^{-1}$. For dividends paid at a
continuous rate $q$, the substitution simply becomes
$$
  S(t) \rightarrow S(t) e^{-q(T-t)}
$$
If dividend payments are known in terms of another
tradable, e.g. a bond, the situation becomes more complicated.
This is so because a dividend payment of $\d_i$ units
of a tradable $P$ at time $t_i$ has the effect of
reducing the stock-price by a factor
$$
  (1+\d_i \frac{P(t_i)}{S(t_i)})^{-1}
$$
This makes the correction factor on $S$ path-dependent
in general. We will return to this problem in
Ref.~\cite{HooglandNeumann99b}.

\subsection{Recovering Black-Scholes}
\label{sec:black-scholes}
In subsection~\ref{sec:deriving-basic-pde} we derived a very general PDE
for the pricing of contingent claims, when the stochastic terms are
driven by Brownian motion. In this section we show that it reduces to
the standard Black-Scholes equation when the underlying tradables are
log-normally distributed with constant drift and volatilities.  In the
Black-Scholes world, we have a number of stocks $S_i$ with SDE's
$$
\frac{d S_i}{S_i}
= \a_i dt+\s_i\cdot dW(t)
$$
Furthermore we have a deterministic bond $P$, satisfying
$$
P(t,T) = \exp(r(t-T))
$$
or in terms of its differential equation
$$
\frac{d P(t,T)}{P(t,T)}=r dt
$$
with $P(T,T)=1$.  For simplicity we take the interest rate and
volatilities to be time-independent. It is not too difficult to extend
the present discussion to the time-dependent case. In fact the
solution was already computed in the previous section. Our basic
equation, Eq.~\ref{eq:5}, gives for the price of a claim
$$
V_t+\frac{1}{2}\s_i\cdot\s_j S_iS_j V_{S_iS_j}=0
$$
Note that $V$ is explicitly a function of $P$. In the Black-Scholes
formulation it is usually defined implicitly. This can be done by
defining
\begin{equation} 
  \begin{split}
    E(S,t)&=V(P,S,t) \\
    V(1,S,t)&=\frac{E(P(t)S,t)}{P(t)}
  \end{split} 
\end{equation}
Thus we find, as promised,
\begin{equation}
  E_t+rS_iE_{S_i}+\frac{1}{2}\s_i\cdot\s_jS_iS_jE_{S_iS_j}-rE=0
\end{equation}
Let us now consider a simple one-dimensional example, a European
call option. The solution can be easily found using the results of the
previous section. 
\begin{eqnarray*}
  V &=&\int (S(t)\phi(z-\s\sqrt{T-t})-KP(t,T)\phi(z))^+ dz\\
  &=& S(t)\Phi(d_1)-KP(t,T)\Phi(d_2)
\end{eqnarray*}
with
$$
d_{1,2}=\frac{\log \frac{S(t)}{K P(t,T)}\pm \half\s^2(T-t)}{\s\sqrt{T-t}}
$$
This is the well-known Merton's formula~\cite{GemanElkarouiRochet95}.
The homogeneity relation, Eq.~\ref{eq:3}, can be used to derive
relations between the greeks. In this present case it is given by
$$
V=SV_S+PV_P
$$
Indeed, using $V_S=\Phi(d_1)$ and $V_P=-K\Phi(d_2)$, the equality
follows. Since in the Black-Scholes universe $P$ is a deterministic
function of $r$, we have for $\r\equiv V_r$
$$
\r=V_PP_r=-(T-t)PV_P=(T-t)(SV_S-V)
$$
These type of relations were al\-ready obser\-ved in a dif\-ferent con
-text in Ref.~\cite{Carr93}.  Furthermore, Eq.~\ref{eq:10} gives the
following relations
$$
SV_{SS}+PV_{PS}=SV_{SP}+PV_{PP}=0
$$
Again this is easily checked by substitution of the solution $V$.

\subsection{Quantos}
\label{sec:quantos}

Quantos are instruments which have a payoff specified in one currency
and pay out in another currency. The pricing of these instruments
becomes trivial, when we consider the problem using only tradables in
one economy. This requires the introduction of an exchange-rate to
relate the instruments denominated in one currency to ones denominated
in another currency. The exchange-rate is assumed to follow some
stochastic process. In the following we will use a Brownian motion
setting. Let us denote the exchange-rate to convert currency 2 into
currency 1 by $C_{12}$, satisfying
$$
\frac{d C_{12}}{C_{12}}
=\a_{12}dt+\s_{12}\cdot dW(t)
$$
The exchange-rate $C_{21}=C_{12}^{-1}$ to convert currency 1 into
currency 2 then satisfies
$$
\frac{d C_{21}}{C_{21}}
=(-\a_{12}+\s_{12}^2)dt-\s_{12}\cdot dW(t)
$$
Let us consider two assets, one denominated in currency 1, the other
in currency 2, with the following dynamics respectively, ($i=1,2$),
$$
\frac{dx_i}{x_i}=\a_i dt +\s_i\cdot dW(t)
$$
To be able to price the instrument we need two tradables denominated
in one currency. Let us define the converted prices $\tx_1=C_{21}x_1$
and $\tx_2=C_{12}x_2$. The converted prices give us our pairs of
tradables $x_1,\tx_2$ and $\tx_1,x_2$ needed to price the instrument.
The price is identical whether we work in terms of currency 1 or 2.
This is a direct consequence of the scale-invariance of the problem.
For consider first the case where everything is denoted in terms of
currency 1. Then we arrive at the following two SDE's
\begin{eqnarray*}
  \frac{dx_1}{x_1} &=& \a_1 dt +\s_1\cdot dW(t) \\
  \frac{d\tx_2}{\tx_2} &=& (\a_2+\a_{12}+\half\s_2\s_{12})
  dt + (\s_2+\s_{12})\cdot dW(t)
\end{eqnarray*}
Thus the volatilities entering in the pricing problem are $\s_1$
and $\ts_2\equiv\s_2+\s_{12}$. Next consider the case where we
denominate everything in terms of currency 2. The SDE's become
\begin{eqnarray*}
  \frac{d\tx_1}{\tx_1} &=& (\a_1-\a_{12}+\s_{12}^2-\half\s_1\s_{12})
  dt + (\s_1-\s_{12})\cdot dW(t) \\
  \frac{dx_2}{x_2} &=& \a_2 dt +\s_2\cdot dW(t) 
\end{eqnarray*}
In this case, the volatilities which are relevant for the pricing
problem are $\s_2$ and $\ts_1\equiv\s_1-\s_{12}$. Therefore we see
that the difference between calculations in the two currencies
amounts to an overall shift in the volatilities by $\s_{12}$.
But we have already seen that solutions of the PDE, Eq.~\ref{eq:5},
are invariant under such a translation. So we obtain a unique
price function.

\subsection{Heath-Jarrow-Morton}
\label{sec:HJM}
Let us consider the Heath-Jarrow-Morton
framework~\cite{HeathJarrowMorton92}. The common approach is to
postulate some forward rate dynamics and from there derive the prices
of discount-bonds and other interest-rate instruments. But it is
well-known that this model can also be formulated in terms of
discount-bond prices \cite{Carverhill95}. Since discount bonds are
tradables, this approach fits directly into our pricing formalism.
Assume the following price process for the bonds\footnote{Here $d_t$
  denotes the stochastic differential w.r.t. $t$.}
$$
\frac{d_tP(t,T)}{P(t,T)}=\a(t,T,P)dt+\s(t,T,P)\cdot dW(t)
$$
As was mentioned before, the drift and volatility functions should be
homogeneous of degree zero in the bond prices in order to have a
well-defined model.  So they can only be functions of ratios of bond
prices.  In fact the precise form of the drift-terms is not of any
importance in deriving the claim-price.

Let us consider as an example the price of an equity option with
stochastic interest rates. We restrict our attention to Gaussian HJM
models. In that case we have a bond satisfying
$$
\frac{d_tP(t,T)}{P(t,T)}=\a(t,T)dt+\s(t,T)\cdot dW(t)
$$
So the drift and volatility only depend on $t$ and $T$. Note that this
form includes both the Vasicek and the Ho-Lee model. As usual, the
stock satisfies
$$
\frac{dS}{S}=\a dt+\s\cdot dW(t)
$$
Now choosing $P(t,T)$ as a numeraire, we find the following PDE for
the price of a claim (cf. Eq.~\ref{eq:11})
$$
V_t+\frac{1}{2}|\s-\s(t,T)|^2S^2V_{SS}=0
$$
The $|v|$ denotes the length of the vector $v$.  Using the standard
techniques, this leads to the following price for a call option with
maturity $T$ and strike $K$
$$
V(S,P,t)=S(t)\Phi(d_1)+KP(t,T)\Phi(d_2)
$$
with
$$
d_{1,2}=\frac{\log\frac{S(t)}{KP(t,T)}\pm\frac{1}{2}\Sigma}{\sqrt{\Sigma}},
\hspace{5mm} \Sigma=\int_t^T |\s-\s(u,T)|^2 du
$$
Remember that both $\s$ and $\s(t,T)$ are understood to be vectors.
Note that in our model it is not necessary to use discount-bonds as
fundamental tradables to model the interest rate market.  One could
equally well use other tradables such as coupon-bonds or swaps, being
linear combinations of discount-bonds, or even caplets and swaptions.
In our view, it seems to be less natural to model the LIBOR-rate
directly, since this is not a traded object.  In fact,
$\d$-LIBOR-rates are dimensionless quantities, defined as a quotient
of discount bonds
$$
L(t,T) = \frac{P(t,T)-P(t,T+\d)}{\d P(t,T+\d)}
$$
In this res\-pect, the name `LI\-BOR
mar\-ket-mo\-del'\cite{Jamshidian97} seems a contradiction in terms.

\subsection{A trigger swap}
\label{sec:trigger-swap}

Let us now consider a somewhat more complicated example,
a trigger swap. This contract depends on four tradables
$S_i$, and it is defined by its payoff function at maturity $T$
$$
  f(S)=(S_3-S_4){\bf 1}_{S_1>S_2}
$$
Note that both exchange options and binary options are
special cases of this trigger swap. The former is found by
setting $S_3=S_1$ and $S_4=S_2$, the latter by setting
$S_3=P(t,T)$ and $S_4=0$. Let us assume that the $S_i$ satisfy
$$
  \frac{dS_i}{S_i} = \a_i(t)dt+\s_i(t)\cdot dW(t)
$$
For this log-normal model, we can immediately write down the
following formula for the price of the claim
$$
V=\int_{S_1\phi(z-\theta_1)>S_2\phi(z-\theta_2)}
  (S_3\phi(z-\theta_3)-S_4\phi(z-\theta_4))dz
$$
Here, the $\theta_i$ are given by a Cholesky decomposition
of the integrated covariance matrix
$$
\S_{ij}=\int_t^T \s_{i}(u)\cdot\s_{j}(u)du = \theta_i\cdot \theta_j
$$
We will omit the details of the evaluation of this integral.
It is a straightforward application of the procedure described in
subsection~\ref{sec:gener-solut-logn}. The result can be written as
$$
V=S_3\Phi(d_3)-S_4\Phi(d_4)
$$
where
$$
d_i=\frac{\log\frac{S_1}{S_2}+\frac{1}{2}(\S_{22}-\S_{11})
+\S_{1i}-\S_{2i}}{\sqrt{\S_{11}-2\S_{12}+\S_{22}}}
$$
The reader can check that this result is again independent
under gauge-transformations $\sigma_i\rightarrow\sigma_i-\l$,
as it should be. Note that $V_{S_1}$ and $V_{S_2}$ are not in 
general equal to zero. This means that one needs a portfolio
consisting of all four underlyings to hedge this claim. Now let
us consider the special case of an exchange option, setting
$S_3=S_1$ and $S_4=S_2$. In this case, the formulae reduce to
$$
V=S_1\Phi(d_1)-S_2\Phi(d_2)
$$
where
$$
d_{1,2}=\frac{\log\frac{S_1}{S_2}
\pm\frac{1}{2}(\S_{11}-2\S_{12}+\S_{22})}
{\sqrt{\S_{11}-2\S_{12}+\S_{22}}}
$$
In Ref.~\cite{LiuWang99} it is claimed that the value of an
option to exchange two stocks has a dependence on the
interest-rate term structure, or in other words, a dependence
on bond-prices. It should be clear from the discussion above
that this is in fact impossible, because neither the payoff,
nor the volatility functions make any reference to bonds.
Therefore, the price of such an exchange option can be calculated
in a market where bonds do not even exist.

\section{Conclusions and outlook}
\label{sec:conclusions-outlook}
In the preceding sections we have clearly shown the advantages of a
model formulated in terms of tradables only.  In this formulation, the
relativity of prices manifests itself as a homogeneity
condition on the price of any contingent claim, and this fact can be
exploited to bypass the usual martingale construction for the
replicating trading-strategy. The result is a
transparent general framework for the pricing of derivatives.

In this article we have restricted our attention to the problem of
pricing European path-independent claims. The generalization to
path-dependent and American options is straightforward and will be
dealt with in other publications.

Obviously, the applicability of the scaling laws is not restricted to
models with Brownian driving factors. Currently we are considering
alternative driving factors such as Poisson and Levy 
processes. We are also looking at implications for modeling incomplete
markets.  Finally the scaling-symmetry should also hold in markets
with friction. This may serve as an extra guidance in the modeling of
transaction-costs and restrictions on short-selling.

\section{Stochastic differential equations}
\label{sec:stoch-diff-equat}
We use stochastic differential equations to model the dynamics of the
prices $x_\m(t)$ of tradables. The governing equation is given by 
$$
d_t x_\m(t)=\a_\m(x,t)x_\m(t)dt+\s_\m(x,t)x_\m(t)\cdot dW(t)
$$
with initial conditions $x_\m(t)$ and $dW(t)$ denote 
$k$-dimensional Brownian motion with respect to some measure. The
drifts $\a_\m(x,t)$ and volatilities $\s_\m(x,t)$ are assumed to be
adapted to $x$ and predictable. For this equation to have a unique
solution, we have to require some regularity-conditions on the drift
$\a_\m(x,t)$ and volatility $\s_\m(x,t)$. These can stated as
follows~\cite{Gardiner85,Arnold74,BorodinSalminen96}.
\begin{itemize}
\item Lipschitz condition: there exists a $K>0$ such that for all
  $x,y$ and $s\in[t,T]$
  $$
  |\a_\m(x,s)-\a_\m(y,s)|+|\s_\m(x,s)-\s_\m(y,s)|\le K|x-y|
  $$
\item Growth condition: there exists a $K$ such that for all
  $s\in[t,T]$
  $$ 
  |\a_\m(x,s)|^2+|\s_\m(x,t)|^2 \le K^2(1+|x|^2)
  $$
\end{itemize}
The Lipschitz condition above is global, it can in fact be weakened to
a local version.  If the growth condition is not satisfied, the
solution may still exist up to some time $t'$, where the solution
$x_\m(t)$ has a singularity and thus `explodes'.

\bibliographystyle{alpha} 
\bibliography{relativity}

\end{document}